\documentclass[manuscript]{acmart}
\pdfoutput=1 

\renewcommand\footnotetextcopyrightpermission[1]{} 
\usepackage[final]{pdfpages}
\AtBeginDocument{%
  \providecommand\BibTeX{{%
    \normalfont B\kern-0.5em{\scshape i\kern-0.25em b}\kern-0.8em\TeX}}}

\usepackage{xcolor}

\setcopyright{acmlicensed}
\acmJournal{PACMHCI}
\acmYear{2021} \acmVolume{5} \acmNumber{CSCW1} \acmArticle{106} \acmMonth{4} \acmPrice{15.00}\acmDOI{10.1145/3449180}

\begin{document}

\title{Conceptualising Contestability: Perspectives on Contesting Algorithmic Decisions}

\author{Henrietta Lyons}
\email{hlyons@student.unimelb.edu.au}
\orcid{1234-5678-9012}
\affiliation{%
  \institution{The University of Melbourne}
  \streetaddress{Parkville Campus}
  \city{Melbourne}
  \state{Victoria}
  \postcode{3010}
  \country{Australia}
}

\author{Eduardo Velloso}
\email{eduardo.velloso@unimelb.edu.au}
\orcid{0003-4414-2249}
\affiliation{%
  \institution{The University of Melbourne}
  \streetaddress{Parkville Campus}
  \city{Melbourne}
  \state{Victoria}
  \postcode{3010}
  \country{Australia}
}

\author{Tim Miller}
\email{tmiller@unimelb.edu.au}
\orcid{0003-4908-6063}
\affiliation{%
  \institution{The University of Melbourne}
  \streetaddress{Parkville Campus}
  \city{Melbourne}
  \state{Victoria}
  \postcode{3010}
  \country{Australia}
}

\renewcommand{\shortauthors}{Henrietta Lyons, Eduardo Velloso, \& Tim Miller}

\begin{abstract}
  As the use of algorithmic systems in high-stakes decision-making increases, the ability to contest algorithmic decisions is being recognised as an important safeguard for individuals. Yet, there is little guidance on what `contestability'--the ability to contest decisions--in relation to algorithmic decision-making requires. Recent research presents different conceptualisations of  contestability in algorithmic decision-making. We contribute to this growing body of work by describing and analysing the perspectives of people and organisations who made submissions in response to Australia's proposed `AI Ethics Framework', the first framework of its kind to include `contestability' as a core ethical principle. Our findings reveal that while the nature of contestability is disputed, it is seen as a way to protect individuals, and it resembles contestability in relation to human decision-making. We reflect on and discuss the implications of these findings. 
\end{abstract}

\ccsdesc[500]{Human-centered computing~Human computer interaction (HCI)}

\keywords{contestability; algorithmic decision-making; artificial intelligence; algorithmic fairness, accountability, and transparency}

\maketitle

\section{Introduction}

Algorithmic systems, including machine learning and artificial intelligence (AI) systems, are increasingly being used in high consequence decision-making including sentencing \cite{Angwin}, hiring \cite{Melder}, performance evaluation \cite{Houston}, and loan applications \cite{Kharif}. In many cases, these systems are trained on data about past decisions and employ machine learning to produce a prediction, or `output' \cite{Yeung19, Mulligan19, citron2014scored}. In some instances decisions are made autonomously, while in others, outputs are used to assist humans to make decisions. Algorithmic systems have the potential to enhance decision-making by making it more efficient, scalable, consistent, objective, accurate, and cost effective \cite{Lee19, Dignum}. Yet, the use of algorithmic systems in decisions that significantly impact lives has also raised concerns relating to fairness and justice \cite{Binns18, Grgic2018Humanperceptions, Dodge19}, human dignity \cite{Jobin19, Kaminski}, and autonomy \cite{Jobin19, Kaminski}.  

Over the past few years there have been high profile examples of algorithmic systems making discriminatory decisions based on race \cite{Angwin}, enhancing inequality \cite{Eubanks}, and being used in a way that violates due process rights \cite{Houston}. Mulligan et al. \cite{Mulligan19} reflect that ``there is a growing sense that our tools, if not checked, will undermine our choices, our values, and our public policies''. There have been calls for algorithmic systems to be designed responsibly \cite{Dignum}, and in a way that ensures that they are fair, accountable, and transparent \cite{Abdul18, Ribeiro16}. Consequently, numerous principles and guidelines for creating responsible, trustworthy, and `ethical AI' have been developed \cite{Jobin19}. One safeguard that is gaining traction within these ethical guidelines is `contestability'---the ability to contest, appeal, or challenge algorithmic decisions \cite{Mulligan19, Jobin19}. For example, the European Union’s `Ethics Guidelines for Trustworthy AI' sets out that AI systems must be developed, deployed, and used in a fair way, which includes the ability to contest decisions made by, or involving, AI \cite{EUGuide}. Australia's AI Ethics Framework includes `Contestability' as one of its eight core principles for designing, developing and implementing ethical AI \cite{DIISwebsite19}. In addition to these voluntary guidelines, Article 22(3) of the European Union's General Data Protection Regulation (GDPR) provides a legal `right to contest' in relation to decisions made using solely automated processes.

Little is known about what contestability in relation to algorithmic decisions entails, and whether the same processes used to contest human decisions, which range from informal complaints mechanisms to formal legal processes, are suitable for algorithmic decision-making. The ethical guidelines do not provide guidance on what contestability is, what the process of contestation should look like, or how algorithmic systems should be designed, developed, or implemented to ensure that decision subjects can contest decisions made using algorithms. Turning high-level principles into implementable actions is a challenging task for those designing and deploying AI systems \cite{mittelstadt2019principles, Mendoza}. Kaminski highlights that a lack of guidance could disadvantage decision subjects---individuals impacted by algorithmic decisions---who are likely to have different interests to the decision maker: ``This raises the question of whether a company whose interests do not always align with its users’ will be capable of providing adequate process and fair results. There is room for substantially more policy development in fleshing out this contestation right'' \cite{Kaminski}.

Emerging research presents different conceptualisations and interpretations of what contestability in algorithmic decision-making requires and how these sociotechnical systems can be designed to support contestability (e.g. \cite{Almada19, Hirsch17, Bayamlioglu}). Our goal is to contribute to this growing body of work by analysing submissions made in response to a discussion paper, titled “Artificial Intelligence: Australia’s Ethics Framework”, released by the Australian Government. To our knowledge, this is the first AI ethics framework of its kind to propose contestability as a core principle. We seek to understand how contestability is conceptualised as well as how the operation of contestability is envisioned by those who provided submissions.

The submission responses provide a broad range of different perspectives, with submissions from the Australian public sector, the private sector (including multinational companies such as Microsoft and Adobe as well as small to medium enterprises), academics, universities, research institutes, community organisations, industry associations, and individuals \cite{DIISwebsite19}. The submissions provide a greater diversity of viewpoints than if we had focused solely on academic literature. Further, many of the submissions are from entities that are likely to implement contestation processes in relation to algorithmic decision-making in the future: there is value in understanding how these organisations view contestability and envision it operating.

Through a thematic analysis of 65 submissions, we found that while there was a lack of consensus on what the true nature of contestability is, there was agreement that contestability is about protecting individuals. Contestability in algorithmic decisions was also seen as requiring the same processes as contestability in human decisions. To operationalise contestability, submissions highlighted the need for decision makers to address a range of policy considerations such as what can be contested, in addition to providing a contestation process that meets certain design requirements such as accessibility. This data helps to map a minimum range of conceptual debates around contestability in this space. We reflect on the submission responses, using them to prompt discussion around a range of challenges and opportunities that designing for contestability in algorithmic decision-making presents.

Contestability in algorithmic decision-making is a complex and nuanced emerging area of research. Through a rich description of how submissions to Australia's AI Ethics Framework conceptualise contestability in algorithmic decision-making and an analysis of the implications of these findings, we map a range of conceptual debates around contestability that can serve as a trigger for discussion.

\section{Related Work}

\subsection{Contesting algorithmic decisions: Process challenges}

In democratic societies, the ability to contest decisions that have a significant impact on a person's life, whether made by a human or algorithm, is an important way to uphold human dignity and autonomy \cite{Yeung19, Almada19, Mendoza, Kaminski, BrennanMarquez19}. However, recent work indicates that people impacted by decisions made using algorithmic systems can be severely limited in their ability to contest these decisions.

There are a number of jurisdictions, including Australia, the United Kingdom and the United States, where government decision-making is regulated. For example, in the United States, state governments must not ``deprive any person of life, liberty, or property, without due process of law'' (U.S. Const. amend. XIV, § 1). `Due process' generally includes the ability to contest a decision, amongst other rights. Despite being constitutionally enshrined, the AI Now Institute \cite{AINow2018, AINow2019} reports numerous violations of due process rights by states using algorithmic decision-making. Breaches that directly impact a person's ability to contest an algorithmic decision include: a failure to provide notice that a decision has been made; inadequate explanation for a decision that has been made; and, insufficient information about how to appeal \cite{AINow2018, AINow2019}.   In these circumstances, a person has grounds to litigate, however barriers such as cost, time, and lack of knowledge of legal rights can stop individuals from appealing decisions in court. These barriers are particularly concerning when algorithmic decision-making disproportionately impacts vulnerable communities \cite{Eubanks}.

Algorithmic decision-making in the private sector is largely unregulated in a direct manner, a notable exception to this is the European Union's General Data Protection Regulation. However, in light of the numerous ethical AI guidelines in current circulation \cite{Jobin19}, there is a growing expectation that companies design and deploy AI systems responsibly. There are also reputational benefits to providing mechanisms for appeal and dispute resolution. Work relating to content moderation on social media platforms highlights some of the difficulties people face when trying to appeal algorithmic decisions in the private sector. Myers West \cite{MyersWest2018} surveyed 519 social media platform users to understand their experience with content moderation systems, which include both human and automated content review and removal. Myers West found that some platforms did not provide a review process at all. Where a review process did exist, many of the 230 users who chose to make an appeal experienced difficulties with the process. For example, users were often provided with no instructions or information about how to initiate an appeal. Complaints were also made about the failure of decision makers to provide a response or resolution after a decision subject lodged a request for review.
 
Our work contributes to this literature by exploring perspectives from a range of sectors to better understand the expectations around how a contestation process in relation to algorithmic decision-making could operate.

\subsection{Different conceptualisations of contestability}

\subsubsection{Interactive algorithmic systems}

Work in human-computer interaction (HCI) has explored the ability of expert users to interact with a system and to correct errors in order to enhance their understanding of the system and to improve its functioning \cite{Gobry1973, Amershi14, VaccaroCSCW19workshop, Cohn03, caruana06}. There has also been work exploring how users can override or correct low impact algorithmic decisions, such as recommendations, by adjusting a rating \cite{Hirsch17}, or providing feedback via user created tags \cite{Amershi14, Vig2011}. Recent work in HCI extends these avenues of research to consider how algorithmic decision-making systems can be designed to enable users to interactively contest significant decisions made by these systems \cite{Hirsch17}. 

Hirsch et al. \cite{Hirsch17} propose that contestability as a design principle is important because it helps to ensure that a person who might be affected by a system outcome has a chance to correct errors and record disagreement with the system. This is particularly important if there is a risk that the output of the system could be used ``as a blunt assessment tool'' \cite{Hirsch17}. Hirsch et al.'s work focuses on the development of a system that evaluates the delivery of psychotherapy by psychotherapists \cite{Hirsch17}. In this work, the `user', who is able to interact with the system, is an expert, but also a decision subject. To design for contestability, Hirsch et al. suggest four `design strategies': 1) increase model accuracy through testing; 2) ensure legibility through the use of explanation, confidence scores, and the ability to trace the output; 3) provision of training for users; 4) ongoing monitoring of the system, to expose any systemic issues \cite{Hirsch17}. 

Mulligan and Kluttz \cite{Mulligan19} propose that designing for contestability can also contribute to the responsible use of algorithmic systems. This is done by encouraging interactivity between expert user and the system, allowing for critique and correction of the system's reasoning as well as promoting a feeling of shared responsibility \cite{Mulligan19}. Based on this approach, contestability can considered to be a governance mechanism \cite{Mulligan19}. 

\subsubsection{The General Data Protection Regulation}
Article 22(1) of the GDPR sets out a right to not be subject to a decision that has been made using solely automated processing if the decision has a legal (or similar) effect. In the case that a person consents to having a solely automated decision made, or such a decision needs to be made in order to enter into a contract (Article 22(2)), Article 22(3) specifically sets out the right to contest such a decision: ``...[T]he data controller shall implement suitable measures to safeguard the data subject's rights and freedoms and legitimate interests, at least the right to obtain human intervention on the part of the controller, to express his or her point of view and to contest the decision''. While this restriction on automated processing is not new, in comparison to its predecessor, Article 15 of the 1995 Data Protective Directive, Article 22(3) provides stronger safeguards for data subjects \cite{Mendoza}. Guidance suggests that the right to contest under the GDPR requires an internal review process conducted by the decision-making entity, but how this review is carried out is not specified \cite{Kaminski, Workingparty29}.

In recent research interpreting Article 22(3), it is suggested that the ability to contest an outcome will be aided by algorithmic systems being designed to support contestability \cite{Almada19, Bayamlioglu, sarra20}. There is acknowledgment that not just the final decision should be rendered contestable, but the entire decision-making process \cite{Bayamlioglu}.  Almada \cite{Almada19} proposes that to ensure data subjects are provided with the rights the GDPR seeks to protect, including self-determination, contestation should not only be considered a post-decision mechanism, such as an appeal process. Rather, algorithmic systems should be designed to be contestable, an approach Almada calls `contestability by design'. This approach---of designing systems to enable and support contestation---goes beyond simply providing contestation processes such as reviews by the decision-making entity or a court; it is `preventative' \cite{Almada19}. Almada suggests that participatory design be employed, where a representative sample of stakeholders partake in the design process, which will help to detect risks in the system and should reduce the need for contestation. Further, information pertaining to the decision and a pathway to contest should be provided, potentially through digital means such as an interface \cite{Almada19}.

\subsubsection{Recourse}

Recent work introduces the need for `recourse' in relation to algorithmic decisions \cite{Ustan19, Venkatasubramanian20}. Ustan et al. \cite{Ustan19} define recourse as ``the ability of a person to change the decision of the model through actionable input variables''. This definition differs from the notion of legal recourse, which focuses on righting a wrong, for example through providing compensation. Ustan et al. \cite{Ustan19} present a toolkit that provides a ``flipset'', which is a ``list of actionable changes that an individual can make to flip the prediction of the classifier''. In effect, a flipset is an explanation. Venkatasubramanian and Alfano \cite{Venkatasubramanian20} note that underlying Ustan et al.'s notion of recourse, ``recourse narrowly conceived'', is the assumption that the decision was correct. Venkatasubramanian and Alfano \cite{Venkatasubramanian20} highlight that a decision may actually be incorrect because it is based on data that is ``faulty or incomplete'', in which case correcting can occur through an ``appeal''. In either case---\textit{appeal} or \textit{recourse narrowly defined}---it is envisioned that a person would receive a flipset that would describe the changes they need to make to alter the decision of the model \cite{Venkatasubramanian20}.

In our view, contestability is a broader concept than recourse and appeal as described by Venkatasubramanian and Alfano \cite{Venkatasubramanian20}. Beyond the need for an explanation, contestability requires avenues through which decision subjects can elect to contest a decision \cite{Almada19}, and processes that could be used to review algorithmic decisions \cite{Almada19, sarra20, malgieri2019, Watcher18, tubella2020, Workingparty29}.

The above literature considers how algorithmic systems can be designed to support contestability and recourse. It also highlights different conceptualisations of contestability. Our work contributes to this literature by analysing empirical data to understand how contestability is conceptualised by those who made submissions in response to the Australian government's proposed AI Ethics Framework.

\section{Method}

Using thematic analysis, we analysed submissions made in response to a discussion paper released by the Australian Government, titled ``Artificial Intelligence: Australia's Ethics Framework'' (which we refer to as `the Discussion Paper' in the remainder of this paper). We explore how the submissions conceptualise contestability in relation to algorithmic decision-making and how they envision contestability operating.

\subsection{Data collection}
We collected data from submissions made in response to the Discussion Paper which was funded by Australia's Department of Industry, Innovation and Science (DIIS) \cite{Dawson19}. The Discussion Paper proposed eight voluntary ethical principles to be taken into account in the design, development, deployment, and operation of AI in Australia (see Appendix). Principle 7 is ``Contestability: When an algorithm impacts a person there must be an efficient process to allow that person to challenge the use or output of the algorithm''. As far as we are aware, Australia's ethics framework is the first to use contestability as a principle in its own right, creating a unique opportunity for examining reactions to it. In addition to the voluntary ethical principles, the Discussion Paper suggests a `toolkit’ of steps that entities using AI can take to action the principles. A tool that is closely related to contestability is `recourse mechanisms’, which is described as: ``Avenues for appeal when an automated decision or the use of an algorithm negatively affects a member of the public'' \cite{Dawson19}.

Feedback on the Discussion Paper was sought between 5 April 2019 and 31 May 2019 via a `Consultation Hub' on the DIIS website. Feedback could be provided by answering discussion questions in a survey (see Appendix for the list of questions) and/or uploading a written submission. More than 130 written submissions were made in response to the Discussion Paper. The submissions were from the Australian public sector, the private sector (including multinational companies such as Microsoft and Adobe as well as small to medium enterprises (SMEs)), academics, universities, research institutes, community organisations, industry associations, and individuals \cite{DIISwebsite19}. Of these submissions, 116 gave permission to be published and have been made available to the public via the DIIS Consultation Hub \cite{DIISwebsite19}. Those making submissions were asked to provide feedback on any or all of the content in the Discussion Paper, including Principle 7 `Contestability'. 

All 116 submissions were read in full by the first author who recorded the terms used that related to contestability or recourse mechanisms. All submissions were then reviewed again for references to the following terms: contest*, appeal, recourse, redress, remed*, compensation, Principle 7. Sixty five of the submissions included one or more of these terms. The sections of these 65 submissions that referred to these terms became the data set for our analysis. 

A breakdown of the sectors from which submissions were made is shown in Figure ~\ref{Submissions-by-author}. A full list of the authors of these submissions is included in the Appendix. While we did explore the data based on sector, there were no discernible trends, with submissions focusing on a wide range of issues and taking different perspectives. Therefore, we chose to analyse the submissions as a whole rather than by sector.

\begin{figure}[h]
  \centering
  \includegraphics[width=\linewidth]{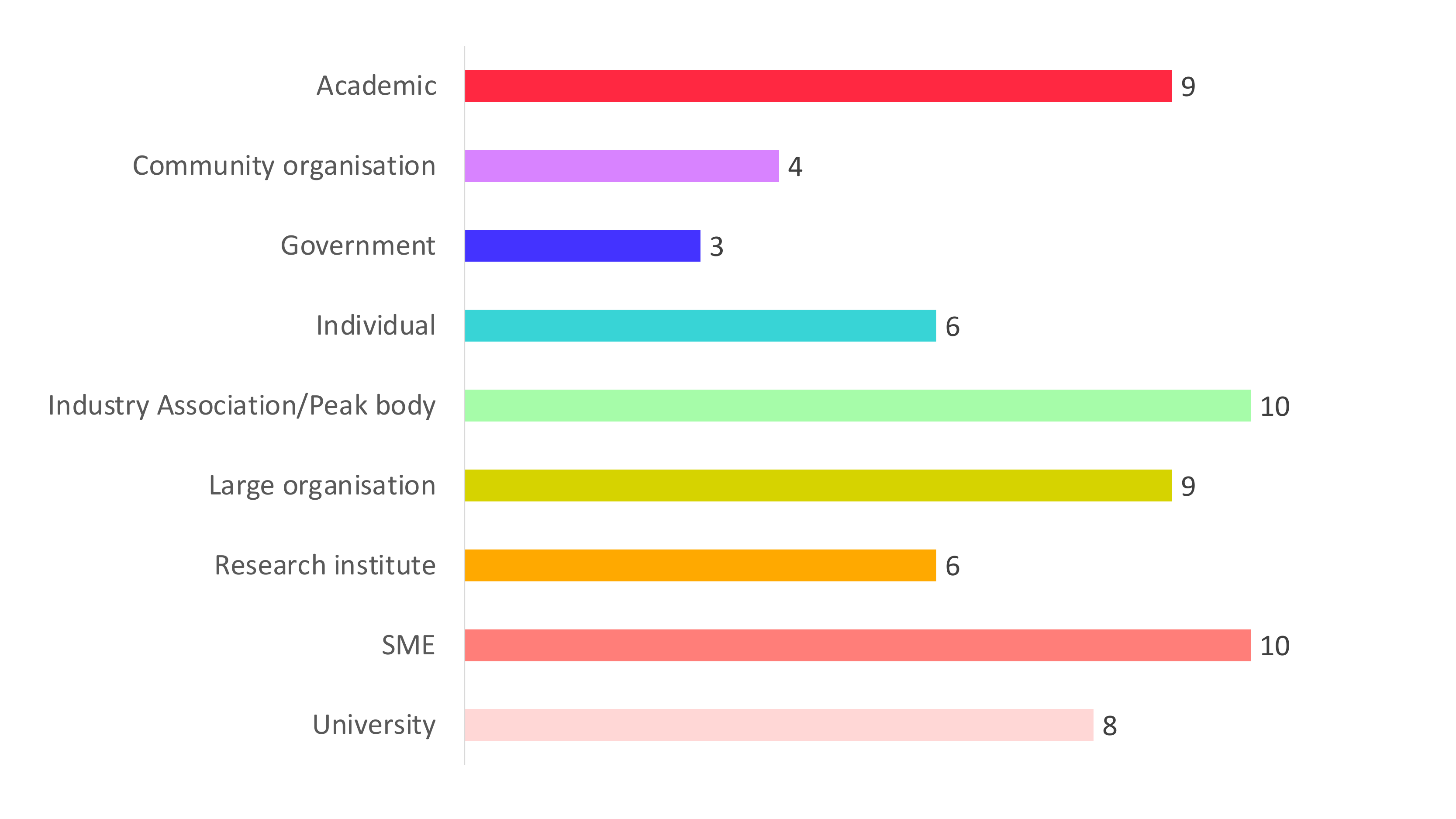}
  \caption{Breakdown of the number of submissions by sector}
  \Description{This figure is a bar chart with the submission sector on the vertical axis and the number of submissions on the horizontal axis. The sectors are: Academic (with 9 submissions); Community organisation (with 4 submissions); Government (with 3 submissions); Individual (with 6 submissions); Industry Association/Peak body (with 10 submissions); Large organisation (with 9 submissions); Research institute (with 6 submissions; SMEs (with 10 submissions); and University (with 8 submissions)}
  \label{Submissions-by-author}
\end{figure}

As the data was collected from publicly available documents informed consent was not sought for this research. The submissions can easily be found using search engines. No login, passwords or registration is required to gain access. There are no statements contained on the website that prohibit the use of these submissions in research.

\subsection{Data analysis}
The first author analysed the data using Braun and Clarke's \cite{Braun} six-stage approach to reflexive thematic analysis, which was chosen for its flexibility and clear procedural steps. Taking a realist approach to the data, the first author inductively coded submissions using NVivo 12. Semantic coding was used to provide a realistic and descriptive account of how contestability is conceptualised. The codes were then sorted into initial themes and all relevant coded data extracts collated into those themes. Themes were reviewed and refined iteratively. Given that this is an under-researched area, we chose to provide a rich thematic description of the entire data set \cite{Braun}.

\subsection{Limitations}
The submissions included in our analysis provide a rich and varied insight into views on contestability. However, it is important to note that the sample is not representative of the general public, with the majority of submissions being from the academic and private sector. While there are a number of submissions from peak bodies, community organisations, and industry associations that advocate on behalf of broader communities, there are many voices that have not been heard in these submissions. In particular, given that algorithmic decision-making has been found to have a disproportionate impact on vulnerable communities \cite{Eubanks}, to avoid perpetuating historical power imbalances it is imperative that members of these communities have their views on contestability taken into account. Future work should explore whether the insights and views presented in the submissions in relation to contestability are aligned to the views of the broader public, and more specifically, to those most likely to be impacted by algorithmic decision-making.

\section{Conceptualising contestability in algorithmic decision-making}

Our first analysis of the submissions in response to the Discussion Paper focused on the research question: \textit{How is contestability in algorithmic decision-making conceptualised?} We address this research question through a qualitative analysis of the submission responses and by reflecting on these responses in the Discussion.

In our qualitative analysis we found that while there was no consensus as to the nature of contestability, with submissions variously viewing it as an end in itself or a means to an end, contestability is seen by the submissions as a way to protect \emph{individuals}. This theme was manifest in talk about the power differential between the decision subject and decision maker, the need for enforceability and redress, and the need for contestation to be a part of a broader regulatory approach. Contestability in relation to algorithmic decision-making was also seen by a majority of submissions as similar to contestability in relation to human decision-making, requiring the same contestation processes, with an emphasis on human review. (Note: because the Discussion Paper, and the submissions in response, refer to ``AI'' rather than ``algorithms'' or ``algorithmic decision-making'' our results and discussion also use this language). 

\subsection{The nature of contestability is disputed}

While contestability is seen as important, there is no consensus on its true nature. Some submissions view it as an ethical principle in its own right; an end in itself. For others, it is a means to an end, either a practical mechanism used to implement other principles or a dimension of higher order ethical principles such as fairness, accountability, or autonomy.

\subsubsection{An end in itself}
 
 Many of the submissions accept the Discussion Paper's inclusion of contestability as an important ethical principle in its own right. While the majority of submissions do not describe why contestability is viewed this way, a couple of submissions do suggest that the ability to contest decisions that have a significant impact on a person's life is a fundamental democratic right.

\begin{quote}
 \textit{``Fundamental to democracy and to a fair society is the principle that power to influence the lives of others should be openly reviewable, transparent, accountable, explicable, and contestable.''} (John Harvey)
\end{quote}

One submission states that contestability is an important principle because it can be used to uncover systems that are not working as they should. This suggests that the ability to contest is seen as a way to regulate the use of AI.

 \begin{quote}
 \textit{``[Contestability] is crucial. Without it, AIs that are failing to achieve what is expected of them can go undetected for long periods of time.''} (The Automated Society Working Group, Monash University)
\end{quote} 

\subsubsection{A means to an end}
Many submissions suggest that contestability is best understood as a means to an end. The submissions provide two different approaches to conceptualising contestability in this way. First, some view contestability as a dimension of other ethical principles that have been included in the Discussion Paper, namely `accountability’ and `fairness’.
 
 \begin{quote}
 \textit{``Transparency, explainability, and contestability are all entailed by accountability, and they all matter precisely to the extent that they contribute to accountability.''} (Australian Academy of Science and the Australian National University) 
\end{quote} 

The second approach to conceptualising contestability as a means to an end is the view that contestability is a tool that supports the practical implementation of other principles and values such as fairness, justice, and autonomy.

\begin{quote}
 \textit{``It is important to recognise that notions of contestability, recourse and redress qualify merely as opportunities that can be given to people so they can exercise their autonomy for their own benefit.''} (Gradient Institute) 
\end{quote} 

\subsection{Contestability is a way to protect individuals}

Decision makers are in a position of power over decision subjects. The use of (often opaque) AI systems in decision-making is seen as widening this power gap. The ability to contest a decision offers decision subjects some protection, allowing them to take back a little control, and to hold decision makers to account. Many submissions highlight the need for contestability to be an enforceable right. The provision of redress is also widely supported, as is the need for additional mechanisms to regulate the use of AI in decision-making.  

\subsubsection{Power differential}

The submissions highlight the difference in power between decision subjects and decision makers. Though this is not a new observation, the opacity of AI decision-making systems, whether because of their complexity or due to proprietary claims, is seen as widening this power gap---decision makers are privy to information that is not available to decision subjects.

\begin{quote}
 \textit{``AI ushers in a new form of digital divide: between the types of actionable `knowledge’ available to those with access to the database and processing power and those without such access.''} (The Automated Society Working Group, Monash University) 
\end{quote} 

Contesting a decision provides decision subjects with a way to hold decision makers to account. Many submissions highlight the need to ensure that the processes to contest a decision are easily accessible and support is available for decision subjects.

\begin{quote}
 \textit{``[E]asy appeal mechanisms are crucial given the disproportionate power imbalance between the developers and implementers of AI systems and those they affect who will remain significantly more vulnerable.''} (Keith Dodds Consulting) 
\end{quote} 

Developers, designers, and those who deploy AI systems often rely on proprietary rights, such as trade secrets, to protect their AI systems from competitors. However, these protections also diminish the ability to scrutinise the systems, leaving decision subjects with very little information on which to base any grounds to contest. A number of submissions suggest that new laws be introduced to limit the use of these proprietary rights in order to provide decisions subjects with more equal footing for contestation.

\subsubsection{Enforceability}

Many submissions call for contestability, and the other core principles, to be made enforceable through legislation. There is concern that if the ethical principles remain voluntary, decision makers may be selective about how and whether they follow them, and decision subjects may not be provided with avenues to contest. Further, there are no consequences for decision makers if they do not follow the guidelines. 

\begin{quote}
 \textit{``The Law Council is concerned that there can be no proper form of contestability without enforceability... Law Council recommends that an ethical and regulatory framework be implemented formally so as to provide for enforceability.''} (Law Council of Australia) 
\end{quote} 

\subsubsection{Redress}

Numerous submissions argue that while the ability to contest a decision is useful, decision subjects will require `redress', `recourse', or `remedy' such as a new decision, compensation for any harm erroneously caused, or reversal of the effects of a decision. To ensure that redress is available, submissions emphasise the importance of enforceability. 

\begin{quote}
 \textit{``While the proposal rightfully stresses the necessity of recourse, it is light on discussing what actual teeth relevant regulation might provide that would be able to give adequate remedy.''} (Otzma Analytics Pty Ltd) 
\end{quote} 

\subsubsection{Part of a broader regulatory approach}

There is acknowledgement in a number of submissions that contestability should not be the only tool used to regulate the use of AI because this would place too much of a burden on decision subjects, who need to initiate contestation. Other mechanisms are needed to provide more fulsome protection to decision subjects.

\begin{quote}
 \textit{``The onus of contest shouldn’t rest solely with victims...We believe that further consideration needs to be given to mechanisms that encourage organisations to proactively identify situations where harm has occurred and then introduce measures to address them as they arise.''} (Future AI Forum (KPMG)) 
\end{quote} 

Further, there are aspects of an AI system that are difficult for individuals to contest, such as the data that was used to train the system. Discrimination and bias are also difficult to perceive at an individual level. The submissions recommend that mechanisms such as freedom of information, independent oversight, ongoing monitoring, and auditing be used in addition to contestability in order to effectively regulate the use of AI. Independent oversight is a particularly popular suggestion, which can be used as a tool to guide the implementation of AI systems and their outputs generally, and also to strengthen contestability.

\begin{quote}
 \textit{``We also suggest adding a ninth principle covering independent oversight. Contestability and accountability require an independent third party to ensure a fair and just outcome.''} (Australian Library and Information Association) 
\end{quote} 

\subsection{Contestability in algorithmic decision-making resembles contestability in human decision-making}
\label{sec:results:akin-to-human-decision-making}

By proposing contestation processes that already exist in relation to human decision-making, submissions indicate that contestability in algorithmic decision-making is envisioned in a way that resembles contestability in human decision-making. 

\subsubsection{Established review mechanisms}

Submissions draw strongly from contestation processes currently used in relation to human decision-making. These review mechanisms can be divided into two broad categories. First, `internal review' where the body responsible for making a decision reviews that decision. Second, `external review', where a separate body to the original decision maker, such as a tribunal, ombudsman, or court, reviews the decision. 

While many submissions raise questions about what a contestation process could look like, in effect seeking clarity from the Australian Government around process requirements, some submissions clearly advocate for specific processes. The Law Council of Australia describes the contestation process that could be used for public decision makers (internal review, followed by independent regulator, then judicial review) and a different process for private sector decisions (internal dispute resolution then referral to an independent regulator). Both of these suggestions reflect processes currently in existence for human decision-making. 

The National Australia Bank goes further than suggesting similar processes, and states that decision-making using AI should not be distinguished from human decision-making and that \textit{``accountability and liability should be applied consistently across both.''} That is, the same protections provided for human decision-making should be provided for decisions made using AI. Similarly, submissions from the medical sector argue that existing processes relating to human decision-making should be used for decisions made using AI. 

\begin{quote}
 \textit{``[T]here are well established processes in medicine for a patient to contest a decision taken about their care, such as via AHPRA or the health ombudsman. These entities are experienced at dealing with the complexities and sensitivities of health. RANZCR believes that these existing processes ought to remain the primary method for resolution for a complaint about medical care.''} (The Royal Australian and New Zealand College of Radiologists) 
\end{quote}

 One submission, from Deakin University, contemplates the potential need for new review mechanisms so that decision subjects can challenge the design of an algorithm.

\begin{quote}
    \textit{``[The ability to challenge the design of the algorithm]...might require new institutional mechanisms to offer consumers, patients, citizens, etc. individual and collective capacity to meaningfully contest the context where AI determines or influences outcomes.''} (Deakin University)
\end{quote}

The AI Now Institute provides one of the few submissions that takes a different view of how contestability can be conceptualised: it contemplates designing due process mechanisms into AI systems, which would help to ensure that decision subjects can contest decisions. In addition, submissions from the health sector contemplate how clinicians (rather than decision subjects) can contest the output of AI systems within the system. This focus on designing systems to support contestability differs from the approach taken by the majority of submissions that envision review processes outside the AI system.

\subsubsection{Human review}

A feature of the processes used to review human decision-making is that the reviewer is always human. If AI can be used to make decisions, there is the possibility that AI can also be used to review or remake decisions. However, a number of submissions explicitly refer to the need for human review. One submission argues that there is an innate right to have a final decision made by a human. 

\begin{quote}
\textit{``All individuals have the right to a final determination made by a person''} (Anand Tamboli, KNEWRON Technologies).
\end{quote}

Academics Matthew Thomas and Katie Miller argue that human review is a necessity even when there is a human-in-the-loop for the decision making: \textit{``[T]he requirements of transparency, explainability and the right to human review should apply whenever AI systems have been used to assist in the decision-making process, even if the final decision is made by a human.''} 

However, submissions also outline challenges associated with human review, including decision fatigue, complacency, bias, and scalability.

\begin{quote}
 \textit{``Contestability exposes decision making processes to undue bias and there can be a lack of reproducibility in the event that the contested decision is remade by a human.''} (National Australia Bank) 
\end{quote}


\section{How contestability in relation to AI could operate}

\begin{figure}[h]
  \centering
  \includegraphics[width=0.8\linewidth]{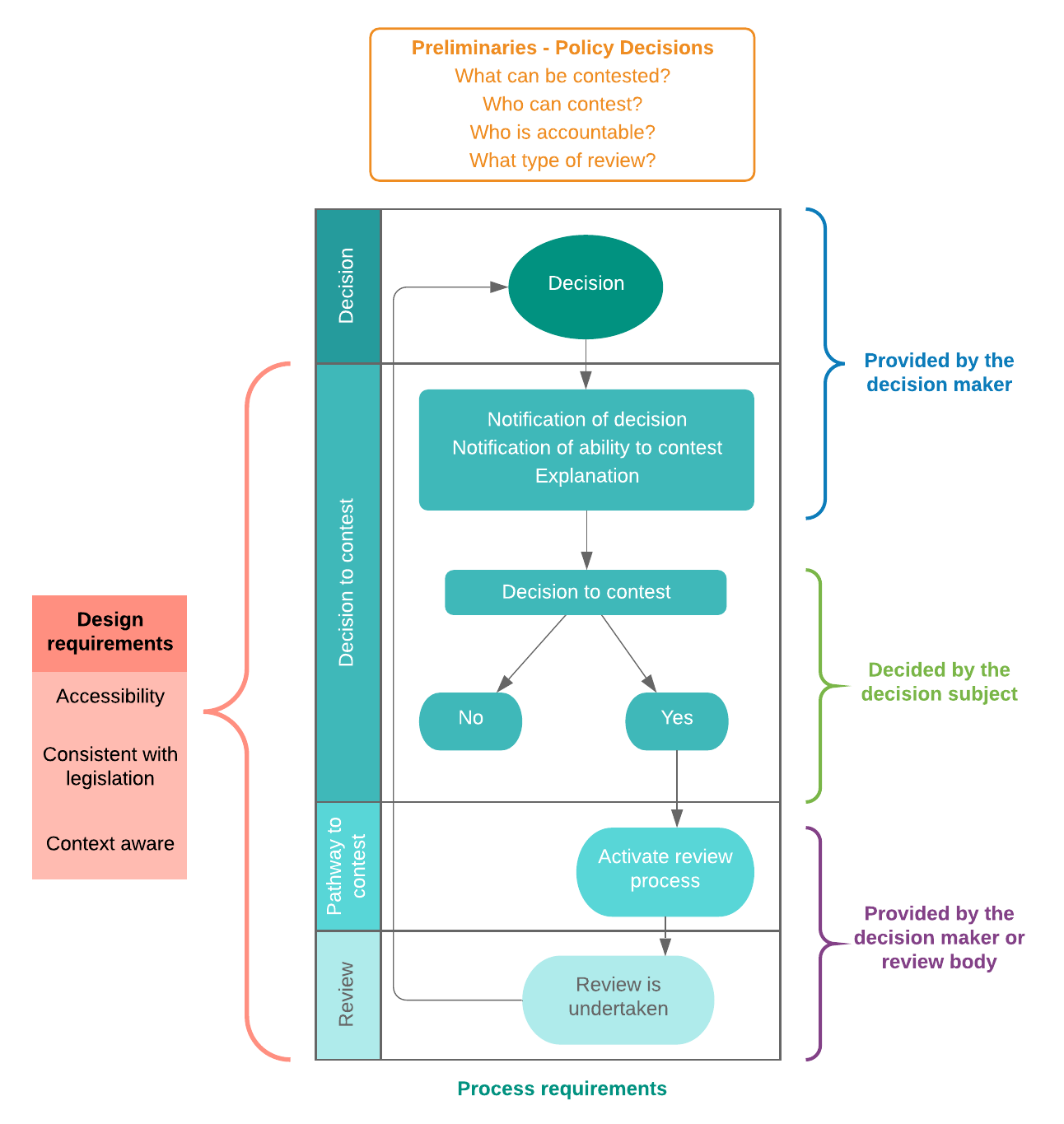}
  \caption{A summary of how contestability could operate based on the  submission responses}
  \Description{This figure summarises the results of the second analysis. The figure contains a flowchart that illustrates the contestation process, which includes: a decision with a downward arrow to a box containing the words `Notification of decision; Notification of ability to contest; and explanation', a downward arrow connects this box to 'decision to contest' which flows to 'yes' and 'no'. The 'yes' box flows to 'Activate review process' which flows down to 'Review is undertaken', which flows back up to 'decision'. Above the contestation process is a box that contains 'Preliminaries - Policy Decisions, and four questions: What can be contested? Who can contest? Who is accountable? What type of review? To the left of the diagram is a box titled 'Design requirements' that contains 'Accessibility; Consistent with legislation; and context aware'}
  \label{Contestation_process}
\end{figure}

The second part of our analysis focused on understanding how submissions envisioned contestability operating. There were three distinct requirements. First, there are policy decisions that need to be determined to provide the scope of contestation. Second, the contestation process requires a number of steps, including the provision of an explanation of the decision and a pathway to request a review. Third, the contestation process should be accessible, context aware, and aligned with relevant legislation. Figure \ref{Contestation_process} illustrates how these themes interact.

\subsection{The preliminaries: policy decisions}
\label{sec:results:the-preliminaries}

The submissions raise a number of questions that relate to policy decisions, which either need to be addressed via government regulation or guidance, or determined by the decision-making entity (in the absence of government direction). These policy decisions include specifications around what can be contested, who is allowed to contest, who is held accountable for decisions, and the type of review that will be provided.

\subsubsection{What can be contested}

The submissions support the need for contestability in relation to algorithmic decisions that have a \textit{significant} impact on a person. The need to define \textit{significant} was raised in many submissions. The Consumer Policy Research Centre points out that this definition may change depending on who is defining it: \textit{``[I]t would be beneficial to provide more information on what constitutes `significant impact’ and who determines that - the consumer, the service, or another entity.''} In addition, KPMG notes different types of impact, including reputational, financial, and physical health. 

The Discussion Paper proposes that a person should be allowed to challenge the ``use or output of the algorithm''. While supporting these inclusions, many submissions advocated that this remit be extended. For example, Microsoft submitted that focusing on `the algorithm' was too narrow, and the focus should be on the `AI system' more generally. Submissions also suggested that the grounds to contest be extended to include: how AI systems are implemented and deployed, including the role of any human decision maker; group and systemic impacts such as discrimination; and, the design of the algorithm, including its features, training data, and decision rules. A number of submissions note the difficulties that individuals will face when trying to contest grounds such as the design of the AI system because of the complexity and opacity of many AI systems.

\subsubsection{Who can contest}

Some people may not have the ability or the resources to contest an algorithmic decision that significantly impacts them. A number of submissions suggest the need for allowing class actions on the basis that an AI system is more likely to impact a person because of their group membership rather than on an individual basis. Class actions are also more financially accessible than individual court cases. Submissions also highlight that the government, legal practitioners, and other groups should be able to contest a decision on an impacted person's behalf.


In their submission, the Office of the Victorian Information Commissioner  goes so far as to suggest that people who may be impacted by an algorithm in the future should have the ability to contest.

\subsubsection{Who is accountable}

Numerous different parties could be held responsible for decisions made using AI, including those deploying the systems, or the developers or designers of the system. A number of submissions suggest that the entities that can influence a decision should be accountable for the decision.

\begin{quote}
 \textit{``Accountability needs to land with the right people or institutions - the people or institutions with the power to do something about the problem, or offer compensation or remediation.''} (Professor Kimberlee Weatherall, Libby Young, Dr. Theresa Dirndorfer Anderson, Associate Professor Julia Powles) 
\end{quote}

\subsubsection{Type of review}

As described in Section~\ref{sec:results:akin-to-human-decision-making},  suggestions for the type of review that should be afforded are heavily influenced by existing processes for reviewing human decision-making including internal (complaints mechanism, internal review) and external review mechanisms (ombudsman, independent regulator, court). Internal review processes are seen as practical because the entity deploying a system is likely to be in the best position to reverse that decision or provide redress. However, Adobe highlights that a requirement for internal review could be burdensome on small businesses, which have less resources to manage review processes compared to large businesses. 

Submissions suggesting `external review' mechanisms, often highlight the importance of `independent' review mechanisms where a different party or organisation to the decision maker reviews the decision. A number of submissions suggest the use of judicial review, which is an independent external review mechanism. However, the following disadvantages of going to court are also highlighted: expense, inefficient processes, reliance on underfunded and overworked legal aid organisations, and increased pressure on a strained system.

Only a couple of submissions consider how a decision could be contested within the AI system itself, or using another AI system, as described in Section~\ref{sec:results:akin-to-human-decision-making}. 

\subsection{Process requirements} 

\subsubsection{Steps in the process}

Submissions describe a number of steps that need to be taken in order to contest a decision. First, a \textit{decision} needs to be made. Second, in order to \textit{decide whether to contest} the decision, a person needs to know that a decision has been made and that there is a contestation process. They will also need some information about the decision---an explanation---to assess whether contestation is required. Once a person has chosen to contest a decision, there needs to be a \textit{pathway to contest} that enables access to the review process. None of the submissions provide detail on what this pathway might look like, for example whether the decision subject could request a review via a computer interface, or whether a more traditional process should be provided, such as lodging a request for review via letter. All that is apparent from the submissions is that whatever the process, it must be clear. Finally, a \textit{review} of the decision will need to be undertaken. As detailed above, in Section~\ref{sec:results:the-preliminaries}, there is no consensus on what a review process should look like, or who should manage this process. Once a review has been completed, a new decision or outcome will be made, which again, will need to be provided to a decision subject, and so the process will continue until there are no further opportunities to contest. 

\subsubsection{Explanation}

The provision of an explanation is not only seen as complementary to contestability, but as a necessary step that enables a person to contest a decision. A number of submissions highlight the link between contestability and explanation. An explanation helps people to understand a decision and allows experts to verify a system's reasoning. Explanations also provide grounds upon which to contest. Many submissions highlight that the opacity, or `black box’ nature, of AI decision making systems is a major challenge for contestability because it makes it difficult to understand why a decision was made. Thus, the submissions see the ability to explain machine decisions---explainability---to be important.  

\begin{quote}
 \textit{``...[I]n order for individuals to contest an algorithm...decision-making processes should be explicable.''} (Office of the Victorian Information Commissioner)
\end{quote}

In terms of what needs to be included in an explanation, the submissions are scant on detail. However, a number of submissions note that an explanation of the decision-making processes should be provided. The submission from Microsoft provides more detail, suggesting the need for different explanations depending on the stakeholder.

\begin{quote}
 \textit{``In terms of transparency, rather than focusing on the algorithm, it would be more useful to pursue a contextual explanation geared to the needs of the particular stakeholders, which may require explanation of one or more of a number of elements of a system. For example, in some cases, a stakeholder may require an explanation of why the resulting model produced a particular output or prediction, or the ways in which the broader system used that output.''} (Microsoft) 
\end{quote}

\subsection{Design requirements}
\label{sec:results:design-requirements}

\subsubsection{Accessibility}

 Submissions emphasise the need for the process to contest to be accessible. The Discussion Paper, in its descriptor for contestability, proposed that the process to contest be `efficient'. Submissions suggest that contestation and redress have the potential to become complex, drawn out processes that could cause further harm to people, so efficient processes are seen as important. 

\begin{quote}
 \textit{``If...there are a flurry of objections raised by concerned individuals, would this stymie the governance chain? Would it block up decision-making processes? Would redressal be long and drawn out causing hardship to the individuals concerned?''} (Values in Defence and Security Technology Group, University of New South Wales) 
\end{quote}

However, many submissions suggest that more than efficiency is required, highlighting the importance of accessibility: the process needs to be clear, easy to access, and affordable. A couple of submissions suggest that support should be provided by the government to decision subjects to access contestation. Having appeal avenues that are easy to access is seen as a way to support decision subjects who lack power in comparison to decision makers. A number of submissions also view accessibility as a way to ensure equality between decision subjects, recognising that some individuals may be less able than others to contest a decision made using AI due to a lack of education or a lack of resources, such as money and time. 

\begin{quote}
 \textit{``Principle 7 on contestability is very necessary...a commitment to facilitating and resourcing the process is crucial to ensure that all individuals have an equal opportunity to benefit from the principle.''} (Wendy Rogers, Macquarie University)
\end{quote} 

Numerous submissions also highlight the need for general education about AI to empower individuals to contest. Again, submissions suggest that it is the role of the government to provide this base level of education. 

\begin{quote}
 \textit{``We recommend a government program to uplift the public understanding of AI to ensure: people understand the concept of a probabilistic system and how it may impact them; people understand their rights in terms of contestability where they are harmed by AI; and people are able to identify when their rights are being infringed upon.''} (KPMG) 
\end{quote}

\subsubsection{Consistent with legal requirements}

In many contexts legislation already exists that provides people who have been affected by decisions with avenues for review. For example, decisions made by the public service that impact individuals is regulated in Australia, and decisions are usually reviewable. Similarly, appeal mechanisms are generally available for decisions made by judges.

Existing contestation processes and legislation will influence the type of review available, and potentially who is accountable for a decision, who can contest, and what can be contested. Contestability is discussed in the submissions in relation to specific areas of law, including administrative law, fair work legislation, consumer protection regulations, medical law, human rights law, and criminal law. Submissions also highlight broader legal principles, such as procedural fairness and due process that apply in a number of contexts. 

\begin{quote}
 \textit{``[I]t is essential that the process for challenging harmful algorithmic impacts not only be efficient but also consistent with broader equitable frameworks such as Due Process.''} (AI Now Institute) 
\end{quote} 

\subsubsection{The context of the initial decision}

Submissions suggest that how the initial decision is made will have an impact on the design of the contestation process. For example, academics, Matthew Thomas and Katie Miller suggest that the power wielded by the government, coupled with an individual's inability to disengage, call for a higher standard to apply to government decisions.

Additionally, when referring to AI decisions some submissions imply automated decisions while others contemplate human-in-the-loop decisions. A couple of submissions suggest that the same protections will be required regardless of how the decision was made.

\begin{quote}
 \textit{``...[W]hat does it matter if the decision was made by a person or an algorithm? It is the basis of the decision that is more important, and the ability to appeal it that is more significant.''} (The University of Queensland, Centre for Policy Futures) 
\end{quote}

\section{Discussion}

The importance of the ability to contest algorithmic decision-making has been highlighted in numerous ethical AI documents as well as in academic research. However, what contestability means in this context and how it should operate is under-researched. With the aim to build a more nuanced understanding of contestability in algorithmic decision-making, we analysed submissions made in response to Australia's AI Ethics Framework Discussion Paper. Our analysis elucidated three key themes in relation to the conceptualisation of contestability: (1) its nature is disputed; (2) it's a way to protect individuals; and (3) it resembles contestability in relation to human decision-making. Interestingly, a majority of the submissions viewed the ability to contest as a post-hoc mechanism such as an appeal process, rather than a mechanism that can be designed into an AI system. In relation to the operationalisation of contestability, submissions outline a need to clarify preliminary policy questions, such as `what can be contested?'; provide a process that includes notification of a decision and an explanation from which grounds to contest can be sourced; and ensure that the process provided is accessible, in line with relevant legislation, and context-aware. In the discussion below, we reflect on these findings, using them as a trigger for highlighting a range of challenges and opportunities that designing for contestability in algorithmic decision-making presents.

\subsection{What is contestability really?}

The lack of consensus as to the nature of contestability is not limited to these submissions, but can be seen in ethical AI guidelines more broadly. In their review of 84 AI ethical guidelines, Jobin et al \cite{Jobin19} cite at least 15 guidelines that classify contestability, redress and remedy variously, sometimes as standalone principles \cite{Torontodec, WEFwhitepaper} and sometimes as mechanisms to action principles including transparency \cite{Uniglobal, ACMstatementontransp, SIIA}, accountability \cite{Internetsociety, Google, EUGuide, ACMstatementontransp, EUrobotics}, empowerment of the individual \cite{ICDPPC}, explicability \cite{SIIA}, and fairness \cite{EUGuide}. 

Perhaps what can be surmised from this lack of consensus is that contestability is important for many reasons: it is a way to hold decision makers to account; it is a way to provide procedural fairness to decision subjects; it upholds the values of dignity and autonomy, and as such, is an important mechanism in democratic societies. Further research could investigate whether targeting one of these goals specifically impacts how the contestation process is designed. 

From a practical perspective, having contestability as a standalone principle highlights its importance and makes it difficult for a decision maker utilising an algorithmic system to selectively choose whether to provide the ability to contest. As a dimension of, or mechanism that supports, other principles, the ability to contest could be replaced by other tools that allow decision makers to satisfy the principles. For example, providing an explanation may satisfy the principle of transparency with contestability becoming an optional requirement. 

\subsection{Problems with seeing contestability as protection}

Noting the power that decision makers wield when they make decisions that significantly impact individuals' lives, many of the submissions view contestability as a way to protect decision subjects. Yet, at the same time, there is acknowledgement that contesting places a burden on the decision subject: the onus is on the individual to pursue an appeal. If viewed solely as a form of protection, we must consider how effective contestation will be. As the Gradient Institute points out in its submission: ``It is important to recognise that notions of contestability, recourse and redress qualify merely as opportunities that can be given to people so they can exercise their autonomy for their own benefit. Some people have less knowledge, resources or time than others and most certainly there will be significant differences in the extent to which people leverage those opportunities.''

Algorithmic decision-making, particularly by governments, often disproportionately impacts vulnerable members of society \cite{Eubanks}. As an example, over a period of four years, Australia used an automated system to detect and automatically raise ``debts'' by matching data relating to welfare payments to tax returns (now colloquially referred to as `robo-debt' in Australia). The key legal issue with the system was that it reversed the onus of proving a debt \cite{carney2019robo}. This onus rests on the government, but the debt letters pressed individuals to prove that they did not owe a debt. Impacted individuals had the ability to seek review of the debt notice through Australia's administrative appeals tribunal. Yet, despite decisions being made in this tribunal to overturn debts (based on systemic issues) \cite{carney2019robo} and the Australian public generally being dissatisfied with its use, the system continued to be used. Only after a class action was launched did the Australian  Government concede that the system was unlawful and agree to refund debt payments to individuals. In this case, the ability to contest individual decisions helped some individuals overturn the decision, but the system was allowed to remain in place, placing financial pressure on thousands of Australia's most vulnerable people, many of whom did not contest the debt notice. 

Contestability can act as a protection, but it should be considered to be the last line of defence. A better approach to protecting individuals is to ensure that from its very conception, a system is designed lawfully and in consultation with those it will impact. Almada's `contestability by design' concept highlights the importance of participatory design and emphasises preventing harm rather than simply redressing it \cite{Almada19}. The submissions also highlight the need for additional mechanisms that provide ``protection'' such as independent oversight, auditing and monitoring of AI systems. It is important that contestability is not used as an excuse by governments and organisations to avoid designing systems with due care: the AI Now Institute \cite{AINow2019} highlights how governments have used the existence of an appeals process ``to resist oversight and reform, claiming that standard appeals are enough to correct any errors.''

While contestation can be a safeguard and a way to hold decision-makers to account, it can also be thought of as a way to exercise autonomy by allowing individuals to challenge decisions that significantly impact them: a right rather than a protection. This framing is consistent with the purpose behind the GDPR's Article 22 and its predecessor Article 15 of the 1995 Data Protective Directive, which limits the use of solely automated decision-making: ``The primary catalyst for Art. 15 was the potential weakening of the ability of persons to exercise influence over decision-making processes that significantly affect them, in light of the growth of automated profiling practices" \cite{Mendoza}.

\subsection{`Humanity-fetishism'? The focus on existing processes and human review}

Our qualitative analysis indicates that contestability in algorithmic decision-making is conceptualised in the same way as contestability in relation to human decisions. Submissions suggest the use of contestation processes that already exist in relation to human decision-making such as complaints mechanisms, ombudsman, administrative review bodies, and courts. Few submissions consider whether existing contestation methods, including human review, will be fit for purpose in relation to decisions made using algorithmic systems. Existing contestation processes have been designed to review human decision-making and to address human errors and biases. More thought needs to be given to whether these existing contestation processes provide the best review mechanisms given the unique challenges algorithmic decision-making faces, such as opacity. 

There are a number of ways that this preference for existing processes can be interpreted. For instance, perhaps these are the best processes for contestation relating to algorithmic decisions. Alternatively, this could be an example of the \textit{mere exposure effect}, which is where people develop a preference for things they are exposed to \cite{Zajonc}. Or potentially submissions are falling back on known options because they cannot imagine other ways of providing for contestability; their thinking is anchored by processes that already exist \cite{Velloso18workshop, Morris14}. 

The submissions also emphasise the need for human review. Providing human review of decisions made using algorithms that significantly impact a person's life, is arguably a way to ``uphold human dignity'' \cite{Almada19, Mendoza, hildebrandt19incomputable}. Human review can be seen as a quality control mechanism: humans are needed to ensure that the algorithmic models being used are functioning properly and have not made any errors, especially because computers cannot (currently) solve problems within their own systems \cite{Almada19, BrennanMarquez19}. Underlying the call of human review may also be a lack of trust in the decisions made using algorithmic systems. Arguably, the submissions display a degree of `humanity-fetishism'; where human decision makers are viewed as unique and special \cite{BrennanMarquez19}. Only a few submissions note that bringing humans into the review process can introduce problems such as human bias, an inability to scale review, and increased review time and cost. 

At a practical level, it is important to consider how human review of algorithmic decisions would work. For example, what would be reviewed? It is clear that the input data could be checked, but what else? The functioning of the algorithm? Or, would a completely new decision be made with an independent system? If a new decision, what decision rules should be followed? Would the human reviewer be able to rely on information taken into account for the algorithmic decision? These are policy decisions that the decision-maker will need to address. 

There is also scope for the use of technology in the review process. Almada suggests that the review process could involve automation, for example by utilising a ``trusted third party algorithm to automatically review a decision'' \cite{Almada19}. This type of review could consider whether decision rules were followed \cite{kroll2016accountable}, while protecting proprietary rights because it does not require `opening the black box'. Alternatively, building on the work investigating contestability as a design objective, algorithmic systems that enable decision subjects to interact with them via an interface could be developed \cite{Hirsch17, Almada19, Mulligan19, Kluttz19}. These technological review processes will likely face the same criticisms raised in relation to decisions made using algorithms in the first instance relating to individualised justice, dignity, autonomy, and fairness.

\subsection{The opacity of decision-making}
The submissions highlight opacity as a key challenge in designing for contestability. Many algorithmic decision-making systems now rely on advanced computational techniques, such as deep learning. The decision rules and inputs of these ``black box'' systems are often obscured due to their technical complexity or are only interpretable by experts, if at all~\cite{Burrell}. Additionally, the workings of algorithmic decision-making systems may be intentionally hidden from view by the developers and those deploying these systems to protect proprietary interests \cite{Burrell}.

Opacity, in its various forms, results in limited information being provided to decision subjects, which makes it difficult to understand why a decision has been made, and consequently, to contest it in any meaningful way \cite{Houston, Burrell, sarra20, Veale18, waldman2019power}. Veale \cite{Veale18} highlights this difficulty: ``A loan applicant denied credit by a credit-scoring [machine learning] algorithm cannot easily understand if her data was wrongly entered, or what she can do to have a greater chance of acceptance in the future, let alone prove the system is illegally discriminating against her (perhaps based on race, sex, or age).'' 
 
Arguably, opacity is not unique to algorithmic decision-making systems, but is also a challenge with human decision-making. As Kaminski \cite{Kaminski} states ``[h]uman decision-making can be deeply, terribly flawed. Human decision makers can be outright discriminatory; can hold deep-seated biases about race, gender, or class; and can exhibit a host of cognitive biases that invisibly influence outcomes''. Zerilli \cite{zerilli2019transparency} also notes that even judges, the ``most esteemed reasoners'', are subject to prejudice and can make poor decisions. However, there has been much effort made to understand how humans make decisions, what might go wrong, and to devise systems to meet the challenges of bias and opacity in human decision-making. For example, significant decisions are often guided by written rules that clearly set out decision criteria; decision-makers are frequently required to attend training sessions to learn about unconscious bias and how to make better decisions; and significant decision-making is often conducted by panels to increase objectivity.

As Waldman \cite{waldman2019power} suggests ```big data'–powered algorithms and machine learning are just as prone to mistakes, biases, and arbitrariness as their human counterparts.'' Yet, people often view technology as more objective and accurate than humans, which ``shields algorithmic systems from critical interrogation'' \cite{waldman2019power}. The impact of this is that people may not even contemplate that a decision is wrong, and so will not contest. In comparison to human decision-making, we are still in the process of uncovering the problems that can be associated with using algorithmic decision-making, like embedded bias \cite{Angwin}, adaptivity \cite{Bayamlioglu}, and spurious correlation \cite{Bayamlioglu, gandy2010}. Further, the ability to use these systems that make decisions at scale presents new issues for contestation and redress. For example, how will humans reviewers keep up with a high demand for review?

\subsection{The relationship between explainability and contestability}

Submissions highlight the need for an explanation to be provided about a decision. Many scholars have also noted the importance of an explanation in determining whether the algorithmic decision was justified (e.g. \cite{Yeung19}), and to provide grounds for review (e.g. \cite{Watcher18, zerilli2019transparency, mittelstadt2019principles}). As the submissions point out, the opacity of many algorithmic systems is a major challenge for understanding how a decision was made. The field of explainable AI (XAI), seeks to distil explanations from these ``black box'' systems. Research in XAI has begun to focus more on human-centred approaches to developing explanations \cite{miller17, miller2019explanation, Abdul18, Wang19, Adadi18}. However, these works have not focused on providing explanations specifically for contestation, with the exception of Wachter et al \cite{Watcher18}. 

There is limited discussion in submissions about what an explanation used for contestation should include. Research is needed to understand explanation requirements in this context. Guidelines would be particularly useful here because there is a difference between providing an explanation in relation to a specific outcome and justifying why that decision was made \cite{binns2018algorithmic}. Further, there may be a difference between what a decision maker and decision subject see as a useful justification for the decision \cite{binns2018algorithmic}. Another unanswered question is whether an explanation produced to initially explain a decision and help a decision subject decide whether to contest is sufficient for the review process, or whether further explanation is required.

Something to be cautious of is relying on the basic and often technical explanations that are currently producible by XAI because these will limit the grounds for contestation \cite{OVICclosertothemachine}. For example, counterfactual explanations are unlikely to provide enough information to be able to contest a decision in court \cite{OVICclosertothemachine}. In addition, the information provided in an explanation, and the way it is provided, can in theory be manipulated to decrease the likelihood that a person will contest a decision \cite{mittelstadt2019principles}. Despite these challenges, we suggest that explainability can be viewed as an enabler for contestability. Contestability requires an explanation, and ensuring that the decisions and outputs of systems are explainable is a step in the right direction.

\subsection{Designing contestation processes: The role of HCI practitioners}
Decision-making occurs in many different contexts, so there will be no one-size fits all contestation process. For example, as the submissions highlight, any existing legal requirements will need to be considered in the design of the process. Further, many of the parameters of the contestation process, such as what can be contested and the type of review that can be offered, are policy questions. These questions need to be answered either by government through the provision of regulation or guidelines, or in the absence of such guidance, by decision makers. So, while our qualitative analysis provides triggers for discussion and a minimum range of conceptual issues to consider, decision-makers will need to design their contestation processes to suit their specific needs and by taking into account the views of the people impacted by the decision. There is a role for HCI practitioners here, to design novel contestation experiences that are informed by human-centred values and practices. The HCI community is well placed to conduct research that can be used by government to inform these policy decisions \cite{woodruffchi18, Lazar16, Thomas17} or to assist decision makers to make these choices \cite{woodruffchi18}.

\section{Conclusion}

Algorithmic decision-making systems are increasingly being used to make significant decisions. The need for these systems to be developed and deployed in an `ethical' and responsible way is pressing. One safeguard being proposed is the ability to contest algorithmic decisions, `contestability'. Our work contributes to the emerging body of work considering contestability by exploring the views of organisations and individuals that made submissions responding to Australia's AI Ethics Framework, the first framework of its kind to propose `contestability' as a core principle. Through a qualitative analysis and a reflective discussion we outline a range of conceptual debates around contestability that can serve as a trigger for further discussion. We reflect on whether contestability offers the protection submissions envision, whether we should view contestability in algorithmic decision-making differently to contestability in human decision-making, and how opacity can hinder a person's ability to contest an algorithmic decision.

\begin{acks}
We would like to thank our reviewers for their valuable feedback on previous versions of this paper. This research was partly funded by Australian Research Council Discovery Grant DP190103414. Henrietta Lyons is supported by the Faculty of Engineering and Information Technology Ingenium scholarship program. Eduardo Velloso is the recipient of an Australian Research Council Discovery Early Career Researcher Award (Project Number: DE180100315) funded by the Australian Government.
\end{acks}

\bibliographystyle{ACM-Reference-Format}
\bibliography{CSCW2020.bib}

\appendix

\section{Appendix}

\subsection{Core Principles for AI \cite{Dawson19}}

\begin{tabular}{p{1.5cm}p{3.5cm}p{7.5cm}}
\toprule
 \multicolumn{3}{c}{\textbf{Core Principles for AI}} \\
  \midrule
Principle 1 & Generates net-benefits & The AI system must generate benefits for people that are greater than the costs. \\
Principle 2 & Do no harm & Civilian AI systems must not be designed to harm or deceive people and should be implemented in ways that minimise any negative outcomes. \\
Principle 3 & Regulatory and legal compliance & The AI system must comply with all relevant international, Australian local, State/Territory and Federal government obligations, regulations and laws. \\
Principle 4 & Privacy protection & Any system, including AI systems, must ensure people's private data is protected and kept confidential plus prevent data breaches which could cause reputational, psychological, financial, professional or other types of harm. \\
Principle 5 & Fairness & The development or the use of the AI system must not result in unfair discrimination against individuals, communities or groups. This requires particular attention to ensure the ``training data'' is free from bias or characteristics which may cause the algorithm to behave unfairly. \\
Principle 6 & Transparency and Explainability & People must be informed when an algorithm is being used that impacts them and they should be provided with information about what information the algorithm uses to make decisions. \\
Principle 7 & Contestability & When an algorithm impacts a person there must be an efficient process to allow that person to challenge the use or output of the algorithm. \\
Principle 8 & Accountability & People and organisations responsible for the creation and implementation of AI algorithms should be identifiable and accountable for the impacts of that algorithm, even if the impacts are unintended. \\
\bottomrule
\end{tabular}

\subsection{Discussion questions posed by Australia's Department of Industry, Innovation and Science}

\begin{tabular}{p{12cm}}
\toprule
 \multicolumn{1}{c}{\textbf{Discussion Questions}} \\
  \midrule
Are the principles put forward in the discussion paper the right ones? Is anything missing? \\
Do the principles put forward in the discussion paper sufficiently reflect the values of the Australian public? \\
As an organisation, if you designed or implemented an AI system based on these principles, would this meet the needs of your customers and/or suppliers? What other principles might be required to meet the needs of your customers and/or suppliers? \\
Would the proposed tools enable you or your organisation to implement the core principles for ethical AI? \\
What other tools or support mechanisms would you need to be able to implement principles for ethical AI? \\
Are there already best-practice models that you know of in related fields that can serve as a template to follow in the practical application of ethical AI? \\
Are there additional ethical issues related to AI that have not been raised in the discussion paper? What are they and why are they important? \\
Do you have any further comments? \\
\bottomrule
\end{tabular}

\subsection{Authors of the submissions analysed}
\begin{tabular}{p{0.45\textwidth}}
\toprule
Adobe Systems	\\
AI Now Institute	\\
Amara Bains, Bains Consulting	\\
Anand Tamboli, KNEWRON Technologies	\\
Artificial Intelligence Collaborative Network	\\
Australasian College of Dermatologists	\\
Australian Academy of Health and Medical Sciences	\\
Australian Academy of Science and the Australian National University	\\
Australian Academy of Technology and Engineering	\\
Australian Industry Group	\\
Australian Information Security Association	\\
Australian Library and Information Association	\\
Baker McKenzie	\\
BSA The Software Alliance	\\
Communications Alliance	\\
Consumer Policy Research Centre	\\
D2D CRC submission	\\
datanomics	\\
Deakin University	\\
Dr Carolyn McKay, University of Sydney Law School	\\
Dr Monika Zalnieriute and Olivia Gould-Fensom	\\
Dr. Jolyon Ford	\\
Electronic Frontiers Australia	\\
Eliot Palmer	\\
Free TV Australia	\\
Future AI Forum (KPMG)	\\
Gradient Institute	\\
John Harvey	\\
Juxi Leitner	\\
Keith Dodds Consulting	\\
KPMG	\\
Law Council of Australia	\\
Matthew Phillipps	\\
Matthew Thomas and Katie Miller	\\
Maurice Blackburn	\\
Microsoft	\\
\bottomrule
\end{tabular}
\begin{tabular}{p{0.37\textwidth}}
\toprule
National Australia Bank	\\
National Health and Medical Research Council	\\
NSW Young Lawyers	\\
Office of the Information Commissioner Queensland	\\
Office of the Victorian Information Commissioner	\\
Otzma Analytics Pty Ltd	\\
Pax Republic	\\
Professionals Australia	\\
Professor Felicity Gerry QC	\\
Professor Kimberlee Weatherall, Libby Young, Dr. Theresa Dirndorfer, Anderson, Associate Professor Julia Powles	\\
Raymond Sheh, Intelligent Robots Group, Curtin University	\\
RMIT University	\\
RMIT University - Graduate School of Business and Law	\\
Rod Jamieson	\\
Roger Clarke Xamax Consultancy	\\
Sarah Kaur Portable	\\
Skanda Kallur, GradientRisk	\\
Stephen Finch	\\
Telstra	\\
The Automated Society Working Group, Monash University	\\
The Pearcey Institute	\\
The Royal Australian and New Zealand College of Radiologists	\\
The University of Queensland, Centre for Policy Futures	\\
ThinkPlace	\\
ThoughtWorks Australia	\\
Toby Lightheart	\\
University of Melbourne	\\
Values in Defence \& Security Technology Group, University of New South Wales	\\
Wendy Rogers, Macquarie University	\\
\bottomrule
\end{tabular}

\end{document}